\documentclass[prl,twocolumn,tightenlines,footinbib]{revtex4-1}
\usepackage{bm,dcolumn,amsmath,graphicx,amsfonts,amssymb}

\usepackage{hyperref} 
\usepackage{xcolor}
\definecolor{cite}{rgb}{0.,0.,0.85}   
\hypersetup{colorlinks,linkcolor={cite},citecolor={cite},urlcolor={cite}}

\renewcommand{\v}[1]{\ensuremath{\boldsymbol{#1}}}		
\newcommand{\bra}[1]{\ensuremath{\langle #1|}}	
\newcommand{\ket}[1]{\ensuremath{|#1\rangle}}	
\newcommand{\braket}[1]{\ensuremath{\langle #1\rangle}}	
\newcommand{\smallspace}{\rule{0pt}{2.5ex}}

\usepackage[normalem]{ulem}
\definecolor{newc}{rgb}{0.,0.6,0.4}

\newcommand{\A}{\ensuremath{\mathcal{A}}} 
\newcommand{\p}{\ensuremath{\pi}}  
\newcommand{\n}{\ensuremath{\nu}}  

\renewcommand{\section}[1]{\vspace{0.85pt}\paragraph*{\textbf{\textit{\small{#1---}}}}}
\renewcommand{\subsection}[1]{\paragraph*{{\textit{\small{#1---}}}}}

\newcommand{\SeeSupplement}{\footnotemark[99]}

\begin{document} 

\footnotetext[99]{See the Supplementary Material for details, which contains tables of hyperfine anomalies and Refs.~\cite{Fella2020,Sunnergren1998,Seelig1998}.}

\title{Nuclear magnetic moments of francium 207--213 from precision hyperfine comparisons}

\author{B.\ M.\ Roberts}\email[]{b.roberts@uq.edu.au}
\author{J.\ S.\ M.\ Ginges}\email[]{j.ginges@uq.edu.au}
\affiliation{School of Mathematics and Physics, The University of Queensland, Brisbane QLD 4072, Australia}
\date{\today}

\begin{abstract}\noindent
We report a fourfold improvement in the determination of nuclear magnetic moments for neutron-deficient francium isotopes 207--213, 
reducing the uncertainties from 2\% for most isotopes to 0.5\%.
These are found by comparing our high-precision calculations of hyperfine structure constants  for the ground states with experimental values.
In particular, we show the importance of a careful modeling of the Bohr-Weisskopf effect, which arises due to the finite nuclear magnetization distribution. 
This effect is particularly large in Fr and until now has not been modeled with sufficiently high accuracy.
An improved understanding of the nuclear magnetic moments and Bohr-Weisskopf effect are crucial for benchmarking the atomic theory required in precision tests of the standard model, in particular atomic parity violation studies, that are underway in francium.
\end{abstract} 

\maketitle


\noindent
Precision investigations of the hyperfine structure of heavy atoms play a critical role in tests of electroweak theory at low-energy, nuclear physics, and quantum electrodynamics~\cite{AtomicReview2017}.
The magnetic hyperfine structure arises due to interactions of atomic electrons with the nuclear magnetic moment.
Comparing calculated and observed values for the hyperfine structure provides the best information about the accuracy of modeled atomic wavefunctions at small radial distances.
This is particularly important for studies of atomic parity violation, which provide powerful tests of physics beyond the standard model~\cite{GingesRev2004,RobertsReview2015}.
Such hyperfine comparisons require accurate knowledge of nuclear magnetic moments; the poorly understood moments for Fr are currently a major limitation.
Accurate moments are also needed in other areas, including tests of quantum electrodynamics (see, e.g., 
resolution of the Bi hyperfine puzzle~\cite{Skripnikov2018}).
Experiments have been proposed~\cite{Behr1993,DzubaFr1995,DzubaPNCsd2001} and are underway at TRIUMF~\cite{Aubin2013,Tandecki2013} to measure parity violation in Fr.
In this atom, due to the higher nuclear charge, the tiny parity-violating effects are enhanced~\cite{Bouchiat1974,*Bouchiat1975} compared to those in Cs, for which the most precise measurement has been performed~\cite{Wieman1997} and a new measurement is in progress~\cite{Toh2019}.

We perform high-precision calculations of the magnetic hyperfine constants $\A$ for the ground states of ${}^{207-213}$Fr.
We examine in detail the effect of the nuclear magnetization distribution, the Bohr-Weisskopf (BW) effect~\cite{Bohr1950,*Bohr1951}.
This is particularly large for the considered Fr isotopes, with relative corrections of 1.3--1.8\% for $s$-states being 6--8 times that of ${}^{133}$Cs, and must be treated appropriately for precision calculations.
While it is standard to model this effect in heavy atoms assuming a spherical nucleus of uniform magnetization, we show this over-estimates the correction by about a factor of two.
Here, we employ a single-particle nuclear model (e.g.,~\cite{Shabaev1994,Shabaev1995,volotka08a}), 
and demonstrate this significantly improves agreement with experiment~\cite{Grossman1999,Zhang2015} for hyperfine anomalies~\cite{Persson2013}.
The difference between the two models amounts to a correction to $\A$ that is much larger than the atomic theory uncertainty; e.g., it is 1.4\% for $^{211}$Fr $s$-states.
The implications for uncertainty analyses are clear: the BW effect {\em must} be modeled accurately for hyperfine comparisons to provide meaningful tests of atomic wavefunctions.
This is imperative for ongoing studies of atomic parity violation~\cite{GingesRev2004}.

We extract improved values of the nuclear moments $\mu$ for ${}^{207-213}$Fr by comparing our calculations of $\A$ with experimental values.
Currently, hyperfine comparisons allow for the most precise determinations of $\mu$ for these isotopes;
experiments with unstable nuclei may provide new avenues in the near future~\cite{Harding2020}.
From an examination of contributions to 
$\A$ for Fr, and for Rb and Cs,
 we conclude that our calculations, and thus the extracted moments, are accurate to at least 0.5\%.
This is up to a fourfold improvement in precision over previous values.

\section{Hyperfine structure calculations}

The relativistic operator for the magnetic hyperfine interaction is:
\begin{equation}
\label{eq:h-hfs}
h_{\rm hfs}
= \alpha\, {\v{\mu}\cdot (\v{r}\times \v{\alpha})} \, F(r) /r^{3} ,
\end{equation}
where $\v{\alpha}$ is a Dirac matrix 
and $\v{\mu}=\mu \v{I}/I $ with $\v{I}$ the nuclear spin (using atomic units $\hbar$\,=\,$|e|$\,=\,$m_e$\,=\,1, $c$\,=\,$1/\alpha$).
$F(r)$ describes the finite nuclear magnetization distribution, and will be discussed in the following section.
Matrix elements of the operator (\ref{eq:h-hfs}) can be expressed as $\A\braket{\v{I}\cdot\v{J}}$, where 
$\v{J}$ is the electron angular momentum.

For the atomic calculations, we employ the all-orders correlation potential method~\cite{Dzuba1987jpbRPA,DzubaCPM1988pla,DzubaCPM1989plaEn,*DzubaCPM1989plaE1}.
This method has been used, e.g., for high-precision calculations of parity violation in Cs~\cite{DzubaCPM1989plaPNC,GingesCs2002,OurCsPNC2012}, and was used recently by us to investigate correlation trends in $\A$ for excited states of Rb, Cs, and Fr~\cite{Grunefeld2019}.
It reproduces the $s$-state energies and lowest $s$-$p$ E1 matrix elements of Fr to $\sim$\,0.1\%~\cite{DzubaPNCsd2001}.

The orbital $\varphi$ and energy $\varepsilon$ for the valence electron are found by solving the single-particle equation:
\begin{equation} \label{eq:CPM}
\big( h_{\rm HF} + V_{\rm B} + \Sigma^{(\infty)} \big) \varphi = \varepsilon \varphi ,
\end{equation}
where
$h_{\rm HF}$\,=\,$c\v{\alpha}\cdot\v{p}$\,+\,$(\beta-1)c^2$\,+\,$V_{\rm nuc}$\,+\,$V_{\rm HF}$, 
$\beta$ is a Dirac matrix,
$V_{\rm  HF}$ is the Hartree-Fock (HF) potential, and $V_{\rm B}$ is the Breit potential (e.g.,~\cite{JohnsonBook2007}). 
For the nuclear potential, $V_{\rm nuc}$, we assume a Fermi charge distribution,
\begin{equation}\label{eq:Fermi}
\rho(r) = {\rho_0}\,{\left(1+\exp[(r-c)/a]\right)^{-1}},
\end{equation}
where $\rho_0$ is a normalization factor, 
$c$ is the half-density radius, and $a$ is defined via $t\equiv 4a\ln3 = 2.3\,{\rm fm}$.

In Eq.~(\ref{eq:CPM}), $\Sigma^{(\infty)}$ is the correlation potential, 
which accounts for the dominating core-valence correlations.
$\Sigma^{(\infty)}$ includes
two effects to all-orders:\ electron-electron screening and hole-particle interaction~\cite{DzubaCPM1989plaEn,*DzubaCPM1989plaE1}.
It is useful to also construct the second-order potential $\Sigma^{(2)}$~\cite{Dzuba1987jpbRPA} to investigate the role of higher-order effects.
To estimate missed correlation effects within $\Sigma$, 
we introduce scaling factors, 
$\Sigma \to \lambda\Sigma$, chosen to reproduce experimental energies (see, e.g., \cite{OurCsPNC2012}).
The accuracy is already very high, so $\lambda\approx1$ (for Fr, $\lambda_s \simeq 0.994$).
To avoid double-counting, all effects must be included prior to the scaling, including the radiative quantum electrodynamics (QED) effects.
We account for these by adding the potential $V_{\rm rad}$~\cite{FlambaumQED2005} into Eq.~(\ref{eq:CPM}).
The QED effects are included via $V_{\rm rad}$ only for the scaling of $\Sigma$.
A different approach is required to include QED effects into the $\A$ calculations;
we take these corrections from Ref.~\cite{Ginges2017} (see also~\cite{Sapirstein2003b,*sapirstein06a}).

Including the hyperfine interaction, the single-particle core orbitals are perturbed: $\phi\to \phi + \delta\phi$ (and $\varepsilon\to\varepsilon+\delta\varepsilon$).
This leads to a perturbation, $\delta V_{\rm hfs}$, to the HF potential known as core polarization.
To find $\delta V_{\rm hfs}$, the equations
\begin{equation}\label{eq:tdhf}
(h_{\rm HF} - \varepsilon_c) \delta\phi_c =-(h_{\rm hfs} + \delta V_{\rm hfs}-\delta\varepsilon_c)\phi_c
\end{equation}
are solved self-consistently for all core orbitals
[when we include the Breit interaction, $V_B$ is added into Eq.~(\ref{eq:tdhf})].
The hyperfine matrix elements for an atom in state $v$ are calculated as
$\bra{\varphi_v} h_{\rm hfs} + \delta V_{\rm hfs} \ket{\varphi_v}$, which includes core polarization to all orders~\cite{DzubaHFS1984,Dzuba1987jpbRPA}. 
We also include small ($\lesssim$\,1\%) correlation corrections 
that cannot be incorporated within the above-mentioned methods -- the structure radiation (SR) and normalization of states (NS)~\cite{Dzuba1987jpbRPA}.

\section{Nuclear magnetization distribution}

The finite nuclear magnetization distribution, described by $F(r)$, gives an important contribution to the hyperfine structure known as the Bohr-Weisskopf (BW) effect~\cite{Bohr1950}.
For heavy atoms, it is standard to model the nucleus as a ball of uniform magnetization, 
with
\begin{equation}\label{eq:Fball}
F_{\rm Ball}(r) = (r/r_N)^3 \quad \text{for $r<r_N$},
\end{equation}
and $F_{\rm Ball}=1$ for $r\geq r_N$, where $r_N = \sqrt{5/3}\, r_{\rm rms}$.

Here, we use a more accurate nuclear single-particle (SP) model, that has been used in studies of QED effects in one- and few-electron ions~\cite{LeBellac1963,Shabaev1994,Shabaev1995,volotka08a}.
For odd isotopes, we take the distribution as presented in Ref.~\cite{volotka08a}:
\begin{equation}\label{eq:F-odd}
F_I(r) = F_{\rm Ball}(r) \big[ 1- \delta F_I \, \ln(r/r_N) \, \Theta(r_N-r) \big],
\end{equation}
which includes the leading nuclear effects, though neglects corrections such as the spin-orbit interaction (see Ref.~\cite{Shabaev1997}).
Here, $\Theta$ is the Heaviside step function, and 
\begin{equation}
\delta F_I =
\begin{cases} 
\dfrac{3(2I-1)}{8(I+1)}\, \dfrac{4(I+1) g_L - g_S}{g_I I}  & I=L+1/2 \\
\dfrac{3(2I+3)}{8(I+1)}\, \dfrac{4I g_L + g_S}{g_I I}   & I=L-1/2,
\end{cases}
\end{equation}
with $I$, $L$, and $S$ respectively being the total, orbital, and spin angular momentum for the unpaired nucleon~\cite{volotka08a},
$g_L = 1(0)$ for a proton(neutron), and $g_I = \mu/(\mu_N I)$ is the nuclear $g$-factor with $\mu_N$ the nuclear magneton.
The effective spin $g$-factor, $g_S$, is determined from the experimental $g_I$ value using the formula:
\begin{equation}\label{eq:gIgSgL}
g_I = \frac{1}{2}\Bigg[g_L+  g_S  + \left(g_L -  g_S \right)\frac{L(L+1) - S(S+1)}{I(I+1)}\Bigg].
\end{equation}

\begin{figure}
\includegraphics[width=0.475\textwidth]{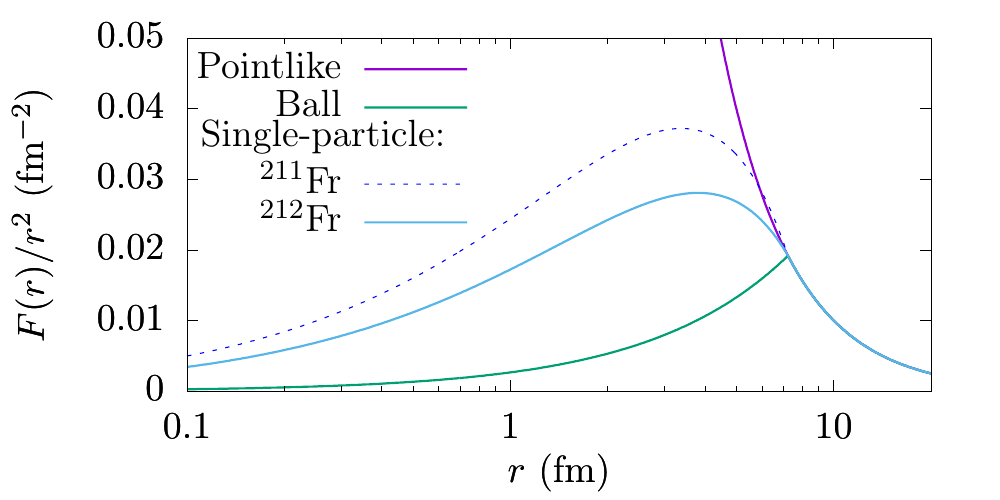}
\caption{Radial dependence of the hyperfine operator~(\ref{eq:h-hfs}), with magnetization distribution as modeled for Fr by:\ a pointlike nucleus, a ball of constant magnetization, and the single-particle model.}
\label{fig:FofR}
\end{figure}

For doubly-odd nuclei with both an unpaired proton and neutron,
the $F(r)$ distribution may be expressed via
\begin{equation}\label{eq:F-doubly-odd}
g_I\,F_I(r) =   \beta\,g_I^{\p} \, F_{I^\p}^\p(r) +(1-\beta)\,g_I^{\n} \, F_{I^\n}^\n(r),
\end{equation}
where $F_{I}^{\p/\n}$ is the unpaired proton/neutron function (\ref{eq:F-odd}),
\begin{equation}\label{eq:beta}
\beta = \frac{1}{2} \left( 1 + \frac{I^{\p}(I^{\p}+1) - I^{\n}(I^{\n}+1)}{I(I+1)}  \right),
\end{equation}
and the total nuclear spin is the sum of that of the unpaired proton and neutron: $\v{I} = \v{I}^\p+\v{I}^\n$.

For the Fr isotope chain between $A$\,=\,$207$--$213$, the proton configuration remains unchanged~\cite{Ekstrom1986},
and the proton $g$-factor for an even nucleus may be taken as that of a neighboring odd nucleus~\cite{Shabaev1995}.
For the unpaired neutron, we determine $g_I^\n$ from the experimental $g_I$ and the assumed $g_I^\p$ and $g_S^\p$ using Eq.~(\ref{eq:gIgSgL}) with $L,S\to I^{\p,\n}$.
The resulting distributions are shown for $^{211,212}$Fr in Fig.~\ref{fig:FofR}.

The relative BW correction, $\epsilon$, is defined via:
\begin{equation}\label{eq:defineEpsilon}
\A_{[F_I]} = \A_{[1]}\left(1 + \epsilon\right),
\end{equation}
where $\A_{[F_I]}$ is calculated using the SP model ($F$\,=\,$F_I$), while $\A_{[1]}$ is calculated assuming a pointlike magnetization distribution ($F$\,=\,1);
both include the finite charge distribution.
Our calculations of $\epsilon$ for $s$ and $p_{1/2}$ states are presented 
in Table~\ref{tab:BW}.
The $\epsilon$ values are stable, depending only weakly on correlation effects~\cite{Ginges2018,Konovalova2018,*Konovalova2020,Barzakh2020}.

\begin{table}
\caption{Literature values for the root-mean-square charge radii ($r_{\rm rms}$), magnetic moments ($\mu$), spin ($I$) and parity ($\Pi$) designations, and configurations for the unpaired proton ($\p$) and neutron ($\n$) for Fr nuclei.
The final columns show the relative BW corrections ($\epsilon$) determined in this work.}
\label{tab:BW}
\begin{ruledtabular}
\begin{tabular}{l D{.}{.}{1.7} D{.}{.}{2.5}  rll  D{.}{.}{3.2}D{.}{.}{3.2}}
$A$
&\multicolumn{1}{c}{$r_{\rm rms}$~\cite{Angeli2013}}
&\multicolumn{1}{c}{$\mu$~\cite{Ekstrom1986}~}
&\multicolumn{3}{c}{Config.~\cite{Ekstrom1986}}
&\multicolumn{2}{c}{$\epsilon$ (\%)}\\
\cline{4-6}\cline{7-8}\smallspace
&\multicolumn{1}{c}{(fm)}
&\multicolumn{1}{c}{($\mu_N$)}
&\multicolumn{1}{c}{$I^\Pi$}
&\multicolumn{1}{c}{$\p$}
&\multicolumn{1}{c}{$\n$}
&\multicolumn{1}{c}{$7s$}
&\multicolumn{1}{c}{$7p_{1/2}$}\\
\hline\smallspace
207	& 5.5720(18)	&3.89(8)	& ${9/2}^{-}$	& $h_{9/2}$	&					&	-1.26	&-0.37\\
208	& 5.5729(18)	&4.75(10)& ${7}^{+}$		& $h_{9/2}$	& $f_{5/2}$ 	&	-1.66	&-0.50\\
209	& 5.5799(18)	&3.95(8)	& ${9/2}^{-}$	& $h_{9/2}$	&					&	-1.29	&-0.38\\
210	& 5.5818(18)	&4.40(9)	& ${6}^{+}$	   	& $h_{9/2}$	& $f_{5/2}$ 	&	-1.67	&-0.50\\
211	& 5.5882(18)	&4.00(8)	& ${9/2}^{-}$	& $h_{9/2}$	&					&	-1.32	&-0.39\\
212	& 5.5915(18)	&4.62(9)	& ${5}^{+}$		& $h_{9/2}$	& $p_{1/2}$	&	-1.77	&-0.53\\
213	& 5.5977(18)	&4.02(8)	& ${9/2}^{-}$	& $h_{9/2}$	&					&	-1.33	&-0.40\\
\end{tabular}
\end{ruledtabular}
\end{table}


To test the accuracy of the nuclear models, we express Eq.~(\ref{eq:defineEpsilon}) as
$\A  = g_I \, {a}^0(1 + \delta)(1 + \epsilon)$~\cite{Persson2013},
where 
$\delta$ is the correction due to the finite nuclear charge distribution.
Here, $a^0$ is the hyperfine constant assuming a pointlike nucleus (for both the magnetization and charge distributions) with $g_I$ factored out.
Importantly, $a^0$ is the same for all isotopes of a given atom~\footnote{We note that we don't directly calculate $a^0$, this is just an instructive factorization.}.

\begin{figure}
\includegraphics[width=0.475\textwidth]{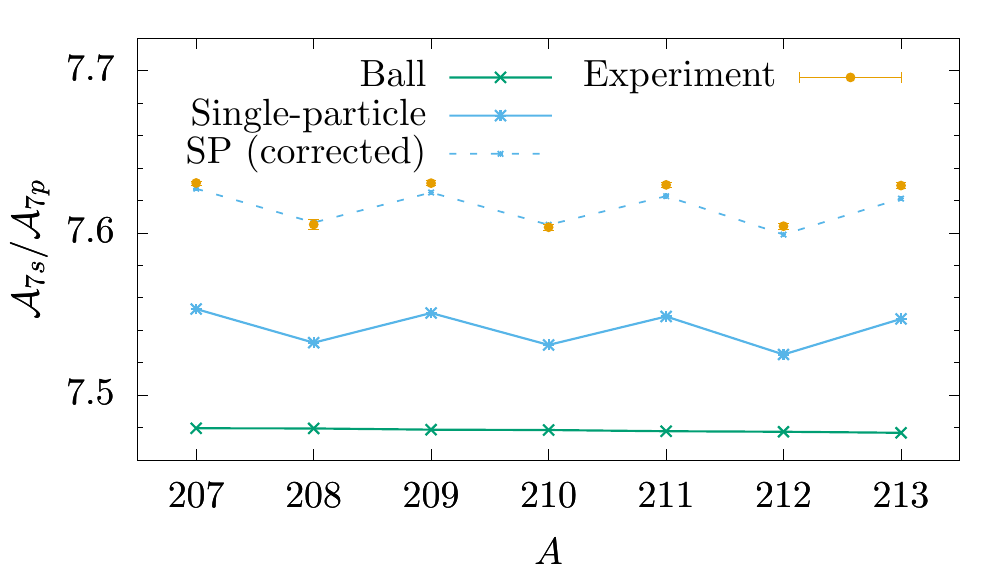}
\caption{Calculated ratios of the $7s$ to $7p_{1/2}$ hyperfine constants for ${}^{207-213}$Fr using the ball and single-particle (SP) nuclear magnetization models, and comparison with experiment~\cite{Zhang2015}.
The odd-even staggering is due to the addition of neutrons; the slight negative slope is due to the changing nuclear radius.
The dashed blue line shows the calculated (SP) ratios corrected by the factor $\xi$($^{133}$Cs); see text for details.}
\label{fig:Asp}
\end{figure}

We form ratios using the $7s$ and $7p_{1/2}$ states for each of the considered Fr isotopes~\cite{Grossman1999} (see also~\cite{Persson2013,Buttgenbach1984,Persson1998}):
\begin{equation}\label{eq:AsAp}
\mathcal{R}_{sp} \equiv {\A_s}/{\A_p} \approx (1 + \epsilon_s -  \epsilon_p + \delta_s - \delta_p)\,{a^0_s}/{a^0_p}.
\end{equation}
The term in the parenthesis (less 1) is the $sp$ hyperfine anomaly~\cite{Persson2013}.
$\mathcal{R}_{sp}$ is independent of the nuclear moments, which for most Fr isotopes are only known to 2\%~\cite{Stone2005}.
A comparison between our calculations and the experimental ratios is presented in Fig.~\ref{fig:Asp}. 
The isotope dependence of $\mathcal{R}_{sp}$ is dominated by $\epsilon_s$ (for Fr, $\epsilon_s>3\epsilon_p$).
Though $|\delta|$\,$>$\,$|\epsilon|$, $\delta$ is modeled accurately by the charge distribution
 (\ref{eq:Fermi}), and changes only slightly between nearby isotopes.
We find errors resulting from uncertainties in $c$ and $t$ (\ref{eq:Fermi}) to be negligible~\SeeSupplement.

Since the proton configuration remains unchanged, the differences in $\mathcal{R}$ along the isotope chain are due to the contribution of the unpaired neutron to the BW effect, $\epsilon^{(\n)}$ (see Fig.~\ref{fig:Asp}).
Thus, we can cleanly extract $\epsilon^{(\n)}$ from the ratio of $\mathcal{R}$ between neighboring isotopes.
Comparing our values to experimental ratios~\cite{Grossman1999,Zhang2015}, we find that we reproduce $\epsilon^{(\n)}$ to between $5$ and $35\%$. 
The neutron contributes about $30\%$ to the total $\epsilon$, see Table~\ref{tab:BW}.

To gauge the accuracy of the calculated proton contribution to $\epsilon$, we consider the well-studied H-like $^{209}$Bi$^{82+}$ ion.
This isotope has the same (magic) number of neutrons as $^{213}$Fr, and the same proton configuration as the considered Fr isotopes.
For H-like ions, the BW effect can be extracted cleanly from experiment, without uncertainties from electron correlations.
Combining high-accuracy calculations including QED effects~\cite{Skripnikov2018,Volotka2012} with a recent measurement of the ground-state hyperfine splitting \cite{Ullmann2017}, the experimental $1s$ BW correction is found to be $-1.03(5)\%$ (see also, e.g.,~\cite{Senkov2002}).
Using the SP model, we calculate the BW correction to be $-1.07\%$, in excellent agreement 
with the experimental value~\SeeSupplement.

As a further test of the proton BW correction, we examine the ratio $\mathcal{R}$.
While $\mathcal{R}$ is independent of the nuclear moments, it depends on electron wavefunctions, and the difference between theory and experiment is likely dominated by errors in the correlations.
We rescale
$\mathcal{R}$ for Fr by the factor
$\xi = \mathcal{R}_{sp}^{\rm Expt.}$($^{133}$Cs)$/\mathcal{R}_{sp}^{\rm Th.}$($^{133}$Cs), 
which empirically corrects the calculated Cs $\mathcal{R}$ value, and amounts to a shift smaller than 1\%.
Since the relative correlation corrections between Cs and Fr are similar~\cite{Grunefeld2019}, this roughly accounts for correlation errors in $\mathcal{R}$(Fr).
After rescaling, we find agreement with experiment to 0.1\% (dashed line in Fig.~\ref{fig:Asp}).
The BW effect contributes about 1\% to $\mathcal{R}$, implying we accurately reproduce the proton contribution to $\approx$\,10\%~\SeeSupplement.

We conclude that the BW effect is calculated accurately, and take the uncertainty to be 20\%.
Note that the BW effect for $^{211}$Fr was calculated recently \cite{Ginges2017} using both Eq.~(\ref{eq:F-odd}) and a more complete model where the nucleon wavefunction is found using a Woods-Saxon potential with spin-orbit effect included.
The extra effects shift $\epsilon$ by $\approx$10\%, well within our assumed uncertainty.

\section{Results and discussion}

In Table~\ref{tab:calculations}, we present our calculated hyperfine constants for $^{87}$Rb, $^{133}$Cs, and $^{211}$Fr, along with experimental values for comparison.
Note that for Fr the uncertainty in the calculated $\A$ is dominated by that of the literature value for $\mu$.
The ratio $\A^{\rm Th.}/\mu$, however, is independent of this uncertainty.

\begin{table}
\caption{Contributions to the ground-state hyperfine constants $\A$ (in MHz) for $^{87}$Rb, $^{133}$Cs, and $^{211}$Fr.
The last two rows show the discrepancy between theory and experiment. 
The Fr calculations assumed $\mu$\,=\,$4.0\mu_N$, which has a 2\% uncertainty \cite{Ekstrom1986}; the resulting 2\% uncertainties for the Fr calculations are shown in italics.}
\label{tab:calculations}
\begin{ruledtabular}
\begin{tabular}{l D{,}{}{3.4} D{,}{}{5.6} D{,}{}{4.8}}
& \multicolumn{1}{l}{${}^{87}$Rb (5$s$)~~~}	
& \multicolumn{1}{l}{$^{133}$Cs (6$s$)~~}	
& \multicolumn{1}{l}{$^{211}$Fr (7$s$)~~}\\
\hline\smallspace
HF	& 2183,.1	& 1433,.7	& 5929,.2\\
$\delta V_{\rm hfs}$	& 460,.3	& 294,.4	& 1111,.2\\
$\Sigma^{(2)}$	& 980,.7	& 779,.6	& 2622,.6\\
$\Sigma^{(\infty)}-\Sigma^{(2)}$	& -171,.1	& -170,.0	& -480,.6\\
$(\lambda-1)\Sigma^{(\infty)}$	& 9,.3	& -5,.1	& -13,.1\\
SR+NS	& -48,.0	& -31,.9	& -129,.0\\
Breit	& 6,.0	& 5,.9	& 33,.0\\
\hline
\smallspace
Subtotal	& 3420,(21)	& 2307,(12)	& 9073,(37)({\it181})\\
\smallspace
BW	& -9,.5(1.9)	& -4,.8(1.0)	& -120,(24)\\
QED~\cite{Ginges2017}	& -8,.3(1.2)	& -8,.8(1.5)	& -55,(12)\\
\hline\smallspace
Total	& 3403,(21)	& 2293,(12)	& 8899,(46)({\it178})\\
Expt.~\cite{Arimondo1977}	& 3417,.341...	& 2298,.157...	& 8713,.9(8)\\
$\Delta$ (MHz)	& -14,.8	& -5,.1	& 185,\,({\it178})\\
$\Delta$ 	& -0,.43\%	& -0,.22\%	& 2,.1\,({\it2.0})\%\\
\end{tabular}
\end{ruledtabular}
\end{table}

To estimate the theoretical uncertainty, we assigned errors individually for each of the important contributions, which are presented in Table~\ref{tab:calculations}.
We take these as twice the difference between the fitted and unfitted all-orders correlation potentials (`$\lambda\Sigma$' row), and $20\%$ for each of the combined structure radiation and normalization of states (SR+NS), Breit, and BW contributions.
We take QED uncertainties of 15--20\% from Ref.~\cite{Ginges2017}.
This leads to theoretical uncertainties of approximately 0.6\%, 0.5\%, and 0.5\%, for Rb, Cs, and Fr, respectively.
We believe these are conservative estimates, justified by the excellent agreement between theory and experiment for Rb (0.4\%) and Cs (0.2\%).
Recent calculations using the same method for $^{135}$Ba$^+$ and $^{225}$Ra$^+$ also have excellent agreement with experiment, with discrepancies of about $0.2\%$~\cite{Ginges2017}.
As a test of the scaling procedure, we perform the calculations using the second-order correlation potential as well.
The unscaled all-orders value, the scaled all-orders value, and the scaled second-order value all agree within about 0.1\%.
Our calculations for Fr are in excellent agreement with those of previous calculations that use a different method (coupled cluster including up to partial triple excitations)~\cite{Gomez2008,Safronova1999}, with deviations of just 0.1--0.2\%, so long as the BW effect, which has been modeled more accurately by us, and the QED corrections, which were neglected in \cite{Gomez2008,Safronova1999}, are accounted for.

By combining our high-precision calculations with the measured $\A$ values, improved values for the Fr nuclear magnetic moments may be deduced as
\begin{equation}
{\mu} = ( {\A^{\rm Expt.}_{7s}}/{\A^{\rm Th.}_{7s}})\widetilde\mu,
\end{equation}
where $\widetilde\mu$ are the values used as inputs in the calculations ($\mu$ in Table~\ref{tab:BW}).
Since the experimental $\A$ values are known to $\lesssim$\,0.01\%, the uncertainty is dominated by the theory. 
The final calculated hyperfine constants for $^{207-213}$Fr and the resulting recommended values for the nuclear moments are presented in Table~\ref{tab:final}.

\begin{table}
\caption{Final theory values for the ground-state hyperfine constant $\A_{\rm 7s}$ for $^{207-213}$Fr, assuming the literature $\mu$ values and theoretical BW corrections presented in Table~\ref{tab:BW},
alongside other values for comparison.
The $\mu$ values from Ref.~\cite{Ekstrom1986} were deduced from measurements made on $^{211}$Fr, and are therefore not independent.
The $^{208}$Fr \cite{Voss2015} value was extracted using $\mu$($^{210}$Fr) \cite{Gomez2008} as reference, so these are also not independent.
The final column shows the recommended $\mu$ values determined in this work.}
\label{tab:final}
\begin{ruledtabular}
\begin{tabular}{l D{,}{}{5.3} l D{(}{(}{4.3} D{.}{.}{2.4} l D{.}{.}{2.4}}
 & \multicolumn{3}{c}{$\A_{7s}$ (MHz)}	& \multicolumn{3}{c}{$\mu/\mu_N$}\\
\cline{2-4}\cline{5-7}
\multicolumn{1}{c}{$A$} & \multicolumn{2}{c}{Expt.}	& \multicolumn{1}{c}{Theory}& \multicolumn{2}{c}{Others}	& \multicolumn{1}{c}{This work}\\
\hline\smallspace
207&	8484,(1)&	\cite{ISOLDE1985}&	8664(45)&	3.89(8)&	\cite{Ekstrom1986}&	3.81(2)\\
208&	6650,.7(8)&	\cite{ISOLDE1985}&	6773(35)&	4.75(10)&	\cite{Ekstrom1986}&	4.67(2)\\
&	6653,.7(4)&	\cite{Voss2015}&	&	4.71(4)\tablenotemark[1]&	\cite{Voss2015}&	\\
209&	8606,.7(9)&	\cite{ISOLDE1985}&	8793(46)&	3.95(8)&	\cite{Ekstrom1986}&	3.87(2)\\
210&	7195,.1(4)&	\cite{ISOLDE1985}&	7317(38)&	4.40(9)&	\cite{Ekstrom1986}&	4.33(2)\\
&	&	&	&	4.38(5)\tablenotemark[1]&	\cite{Gomez2008}&	\\
211&	8713,.9(8)&	\cite{ISOLDE1985}&	8899(46)&	4.00(8)&	\cite{Ekstrom1986}&	3.92(2)\\
212&	9064,.2(2)&	\cite{ISOLDE1985}&	9209(48)&	4.62(9)&	\cite{Ekstrom1986}&	4.55(2)\\
213&	8759,.9(6)&	\cite{ISOLDE1985}&	8943(47)&	4.02(8)&	\cite{Ekstrom1986}&	3.94(2)\\
\end{tabular}
\end{ruledtabular}
\tablenotetext[1]{These values for $^{208}$Fr~\cite{Voss2015} and $^{210}$Fr~\cite{Gomez2008} change to 4.66(4) and 4.33(5), respectively, when corrected to account for the QED and BW effects; see text for details.}
\end{table}

Most of the considered experimental values for $\mu$ come from a single measurement~\cite{Ekstrom1986}.
In that work, the values for $^{207-213}$Fr were deduced from the $^{211}$Fr value.
Our extracted values agree with those values within the uncertainties, though are about 2\% smaller.

A more recent result is available for $^{210}$Fr, which comes from a combination of a measurement and calculation of $\A$ for the excited $9s$ state~\cite{Gomez2008}.
The theory portion of that work used a ball model for the magnetization distribution, and did not include QED effects.
If we re-scale the calculations from Ref.~\cite{Gomez2008} to correct for the BW and QED effects as described above, their value for the $^{210}$Fr magnetic moment changes from $\mu=4.38\,\mu_N$ to $4.36\,\mu_N$ using $\A_{9s}$, or to $4.33\,\mu_N$ using $\A_{7s}$, which are both in agreement with our value.
We note that our calculations~\cite{Grunefeld2019}, as well as those from Refs.~\cite{Gomez2008,Safronova1999}, reproduce the Rb and Cs $\A$ values for the ground states with higher accuracy than for the excited states (see Ref.~\cite{Grunefeld2019}).
Therefore, we expect that it is more accurate to extract $\mu$ using the Fr $7s$ ground state.

A more recent measurement of $\mu(^{208}$Fr) is also available~\cite{Voss2015}.
However, this value and those for $^{204-206}$Fr were found using the $\mu(^{210}$Fr) result of Ref.~\cite{Gomez2008} as reference.
These should therefore be corrected to account for the QED and BW effects.
The corrected result for $\mu$($^{208}$Fr) is 4.66(4)\,$\mu_N$, coinciding with our result. 

\section{Conclusion}

By combining high-precision calculations with measured values for the ground-state magnetic hyperfine constants, we have extracted new values for the nuclear magnetic moments of $^{207-213}$Fr.
In particular, we show the importance of an accurate modeling of the nuclear magnetization distribution, the so-called Bohr-Weisskopf effect, which until now has not been modeled with sufficiently high accuracy for Fr.
We model this effect using a simple nuclear single-particle model, which gives greatly improved agreement for hyperfine anomalies.
We conclude that the single-particle model should be used rather than the ball model in future high-precision calculations.
Our extracted nuclear magnetic moments are about 2\% smaller than existing literature values, which mostly come from a single experiment.
Based on our analysis, we expect our results to be accurate to 0.5\%, a factor of four improvement in precision over previous values for most isotopes.

\acknowledgements
\subsection{Acknowledgments} 
This work was supported by the Australian Government through an Australian Research Council Future Fellowship, Project No.~FT170100452.

~\\
\hrule
~\\

\appendix
\setcounter{equation}{0}
\setcounter{figure}{0}
\setcounter{table}{0}
\renewcommand{\theequation}{S.\arabic{equation}}
\renewcommand{\thefigure}{S.\arabic{figure}}
\renewcommand{\thetable}{S.\arabic{table}}

\section{Supplementary Material}

\noindent
Here, we present details of the calculations used to gauge the accuracy of the calculated Bohr-Weisskopf (BW) corrections for Fr.
As discussed in the main text, Eq.~(\ref{eq:AsAp}) and Fig.~(\ref{fig:Asp}) test both the neutron and proton contributions to the BW effect. Here, we consider these and other tests in more detail.

The differential hyperfine anomaly, ${}^1\Delta^2$, is defined via the ratio 
 for two isotopes (e.g., \cite{Persson2013}):
\begin{equation}\label{eq:anomaly}
\frac{\A^{(1)}}{\A^{(2)}} \equiv \frac{g_I^{(1)}}{g_I^{(2)}}\left(1+{}^1\Delta^2\right).
\end{equation}
For nearby isotopes of the same atom, the correction due to differences in the nuclear charge distributions, $\delta$, will strongly cancel, leading to
\begin{equation}\label{eq:anomaly2}
{}^1\Delta^2  \approx\epsilon^{(1)} - \epsilon^{(2)}.
\end{equation}
The ratio (\ref{eq:anomaly}) is independent of the atomic wavefunctions
though depends on the magnetic moments~\cite{Buttgenbach1984}.

To remove the $\mu$ dependence, one may form~\cite{Persson1998}: 
\begin{equation}\label{eq:doubleAnomaly}
 {}^1\Delta^2_{sp}  
\equiv
\frac{\A_s^{(1)}/\A_s^{(2)}}{\A_p^{(1)} /\A_p^{(2)} } -1 
\approx  \epsilon^{(1)}_s - \epsilon^{(2)}_s - \epsilon^{(1)}_p + \epsilon^{(2)}_p .
\end{equation}
This ``double ratio'' provides a good avenue for testing the nuclear magnetization models, however it depends only on the difference in $\epsilon$ between the pairs of isotopes.
For the Fr isotope chain considered, the proton contributions to the BW effect remain similar and thus cancel in the ratio.
Comparing $ {}^1\Delta^2_{sp} $ for neighboring odd and even isotopes provides an excellent method for testing the accuracy of the neutron BW contribution. 
The double ratios are presented in Table~\ref{tab:A-doubleRatio}.
Comparing our values to experimental ratios~\cite{Grossman1999,Zhang2015}, we find that we reproduce $\epsilon^{(\n)}$ to between $5$ and $35\%$.

\begin{table}[b]
\footnotesize
\caption{\footnotesize Calculated (RPA) and experimental hyperfine ``double'' anomalies [${}^1\Delta^2_{sp}$, Eq.~(\ref{eq:doubleAnomaly})] for several Fr isotopes.
Illustrative theoretical uncertainties are given for the single-particle model, assuming a 15\% uncertainty in the Bohr-Weisskopf correction. The resulting uncertainties are relatively large due to cancellations in $\epsilon_s$ values between isotopes.}
\label{tab:A-doubleRatio}
\begin{ruledtabular}
\begin{tabular}{ll D{.}{.}{3.3} D{.}{.}{2.4} D{.}{.}{2.6}}
\multicolumn{2}{c}{Isotope}	& \multicolumn{3}{c}{${}^1\Delta^2_{sp}$\,(\%)}	 \\
\cline{1-2}\cline{3-5}
\smallspace
1	& 2	& \multicolumn{1}{c}{Ball}& \multicolumn{1}{c}{Single-Particle}	& \multicolumn{1}{c}{Expt.\footnotemark[1]} \\
\hline\smallspace
212	& 207	& -0.03	& -0.41	(8)& -0.349(8) \\
212	& 209	& -0.02	& -0.38(8)	& -0.368(8) \\
212	& 211	& -0.005	& -0.34(7)	& -0.335(9) \\
212	& 213	& 0.01	& -0.32(7)	& -0.356(8) \\
\smallspace
211	& 208	& -0.02	& 0.24(5)	& 0.321(13) \\
211	& 210	& -0.01	& 0.26(6)	& 0.346(10) \\
211	& 212	& 0.005	& 0.35(7)	& 0.336(9)
\end{tabular}
\end{ruledtabular}
\footnotetext[1]{The experimental $\A_{7p_{1/2}}$ values are taken from Ref.~\cite{Zhang2015} for ${}^{207,209,213}$Fr, and from Ref.~\cite{Grossman1999} for the others.
The $\A_{7s}$ values are from Ref.~\cite{ISOLDE1985}, except for ${}^{208}$Fr, where following Ref.~\cite{Zhang2015}, we take a weighted average of those from Refs.~\cite{ISOLDE1985} and \cite{Voss2015}.}
\end{table}

\subsection{Testing the neutron BW contribution}

As an independent test, we consider the BW correction for H-like ions, which may be cleanly extracted from experiment via (see, e.g., Refs.~\cite{Skripnikov2018,Senkov2002}): 
\begin{equation}
\A_0 \, (1+\epsilon) + \delta\A^{\rm QED} = \A^{\rm Expt.},
\end{equation}
where $\A_0$ is the calculated value for the hyperfine constant, including the finite nuclear charge distribution while assuming a pointlike nuclear magnetization distribution, and $\delta\A^{\rm QED}$ is the radiative QED correction.
The well-studied system $^{207}$Pb$^{81+}$ is ideal for testing our modelling of the neutron contribution to the BW effect -- it has a single unpaired neutron, is just one neutron away from being a doubly magic nucleus, 
and has the same neutron structure as $^{212}$Fr.
The result of our single-particle model calculation of the BW effect for $^{207}$Pb$^{81+}$ differs from the experimental value by 8\%,  as shown in Table~\ref{tab:BiPb} (see also, e.g., Ref.~\cite{Senkov2002}).
The ``ball'' model, on the other hand, underestimates the relative BW correction magnitude by close to a factor of 2.

\subsection{Testing the proton BW contribution}

\begin{table}[b]
\footnotesize
\caption{Hyperfine $\A$ constants (in THz) for $1s$ states of H-like Bi and Pb, and relative Bohr-Weisskopf corrections.
$\A_0$ includes the finite nuclear charge effects, though assumes a point-like nuclear magnetization distribution.
We use the most up-to-date values for the magnetic moments, taking 
$\mu(^{208}\rm{Bi})=4.092(2)\,\mu_N$~\cite{Skripnikov2018}
and
$\mu(^{207}\rm{Pb})=0.59102(18)\,\mu_N$~\cite{Fella2020}.
For both cases, the uncertainty in $\A_0$ is dominated by that from $\mu$.
The QED corrections for these ions (and several others) have been calculated a number of times, and are in good agreement with each other; see, e.g., Refs.~\cite{Volotka2012,Shabaev1997,Sunnergren1998,Senkov2002} and references therein.
}
\label{tab:BiPb}
\begin{ruledtabular}
\begin{tabular}{l  D{.}{.}{1.0}l | D{.}{.}{1.1}l}
 {} 	& \multicolumn{2}{c|}{$^{209}$Bi$^{82+}$}& \multicolumn{2}{c}{$^{207}$Pb$^{81+}$}  \\
\hline
\smallspace
$\A_0$			&249.93(12)&	 				& 307.59(9)	 &\\
$\delta\A^{\rm QED}$	&-1.45&\cite{Volotka2012}		&-1.76(2)&\cite{Sunnergren1998}  \\
\smallspace
$\A^{\rm Expt.}$	&245.911(4)&\cite{Ullmann2017}  	&294.00(5)	&\cite{Seelig1998}\\
\hline
\smallspace
&\multicolumn{1}{c}{$\epsilon$\,(\%)}		& 		&\multicolumn{1}{c}{$\epsilon$\,(\%)}		&\\
Expt.				&-1.03(5)		&&	-3.85(4)&\\
This work			& -1.07	&	&-3.55&	\\
\end{tabular}
\end{ruledtabular}
\end{table}

H-like $^{209}$Bi is a particularly good system to use to check the calculations of the proton BW contribution for Fr.
Like the considered odd Fr isotopes, $^{209}$Bi has a single unpaired proton.
It has the same nuclear spin and parity, and a similar magnetic moment, to the considered odd Fr isotopes.
It is only four protons away from Fr, and has the same number of neutrons as $^{213}$Fr (magic number 126).

Our results for H-like Bi are presented in Table~\ref{tab:BiPb}.
We find excellent agreement between our calculated BW correction and the experimental value, as well as with other calculations (see, e.g., Ref.~\cite{Senkov2002}).
In contrast, the ``ball'' model overestimates the relative BW correction magnitude for Bi by a factor of 2.

We also consider tests specific to the case of Fr.
Seeking a measure that is independent of $\mu$ and has sensitivity to the proton contribution to the BW effect,
we form the ratio [Eq.~(12) in the main text]:
\[
\mathcal{R}_{sp} \equiv \frac{\A_s}{\A_p} \approx \frac{a^0_s}{a^0_p}(1 + \epsilon_s -  \epsilon_p + \delta_s - \delta_p).
\]
This depends on electron wavefunctions, and therefore correlations, limiting the ability to accurately extract $\epsilon$.

To circumvent this, we follow Ref.~\cite{Grunefeld2019} and express the total theory value for $\A$ as
\begin{equation}
\A^{\rm Th.} = g_I a^{\rm RPA}(1+\sigma)(1+\epsilon),
\end{equation}
where
$a^{\rm RPA}$ is the RPA (HF with core polarisation) value for $\A$ including the finite-nuclear charge effect (with the $g$-factor factored out), and 
 $\sigma$ is the relative correction due to all many-body effects beyond the RPA approximation.
We may express the ``exact'' value as
\begin{equation}
\A^{\rm Exact} = g_I a^{\rm RPA}(1+\tilde\sigma)(1+\tilde\epsilon),
\end{equation}
where $\tilde\epsilon$ and $\tilde\sigma$ are hypothetical exact values for $\epsilon$ and $\sigma$, respectively.
While the correlation correction $\sigma$ is large compared to $\epsilon$, the {\em errors} in the correlations are of similar magnitude to $\epsilon$.
In the case of Fr, they are likely smaller, due to the large BW effect; the estimated uncertainty coming from correlations for Fr $s$-states is larger than the calculated BW effect (Table II of main text).

We introduce the factor $\xi$, which empirically corrects the calculated Cs $\mathcal{R}$ value:
\begin{align}
\xi^{\rm (Cs)} &= \mathcal{R}_{sp}^{\rm Expt.}(^{133}{\rm Cs})/\mathcal{R}_{sp}^{\rm Th.}(^{133}{\rm Cs}), 
\end{align}
where 
$
\xi \approx 1 + (\tilde\epsilon_s - \epsilon_s) - (\tilde\epsilon_p - \epsilon_p) + \Delta\sigma_s - \Delta\sigma_p,
$
with
$\Delta\sigma \equiv (\tilde\sigma - \sigma)$.
Since $|\epsilon^{({\rm Fr})}|\gg|\epsilon^{({\rm Cs})}|$, this implies: 
\begin{multline}\label{eq:correctedRbyXi}
\xi^{({\rm Cs})} \,\mathcal{R}_{sp}^{({\rm Fr})}  
\approx 
\frac{a_{{\rm Fr},s}^{\rm RPA}}{a^{\rm RPA}_{{\rm Fr},p}}
\Big(1 + \epsilon_s^{\rm (Fr)} -  \epsilon_p^{\rm (Fr)} \\
+ [\sigma_s^{\rm (Fr)} + \Delta\sigma_s^{\rm (Cs)}] - [\sigma_p^{\rm (Fr)} + \Delta\sigma_p^{\rm (Cs)}]\Big).
\end{multline}

\begin{figure}
\includegraphics[width=0.475\textwidth]{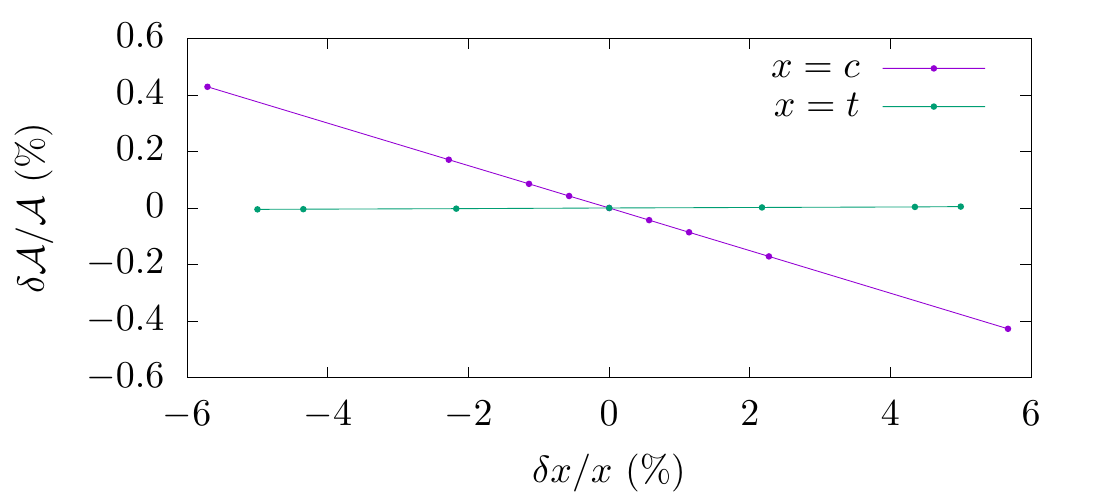}
\caption{Relative sensitivity of the calculated hyperfine structure constant $\A_{7s}$ to the half-density radius $c$, and skin-thickness $t$ (parameters for the Fermi nuclear charge distribution).
The plot is for $^{211}$Fr, though the results are essentially isotope-independent.}
\label{fig:dRrms}
\end{figure}

Since the relative correlation corrections are similar for alkali metals~\cite{Grunefeld2019},
the $\Delta\sigma$ terms are expected to be both small and similar between Cs and Fr.
As such, the same $\Delta\sigma_s^{\rm (Cs)}$ terms should also approximately correct the theoretical Fr $\mathcal{R}$ for missed electron correlation effects.
Though $\Delta\sigma_s$ and $\Delta\sigma_p$ have opposite sign and add in Eq.~(\ref{eq:correctedRbyXi}), multiplying by $\xi$ amounts to a small $<$\,1\% correction.
After the rescaling, we find agreement with experiment to 0.1\%;
a similar result is reached if we instead re-scale by the Rb $\xi$ factor.
Since the BW effect contributes about 1\% to $\mathcal{R}$, this implies we accurately reproduce the proton contribution to the BW effect to about 10\%.
We stress that scaling $\mathcal{R}{\rm (Fr)}$ by the $\xi^{\rm(Cs)}$ factor cannot empirically correct for systematic errors in the BW effects for Cs and Fr, because $\epsilon$ for Fr is an order of magnitude larger than that for Cs.

\subsection{Nuclear charge distribution}

We also quantify possible errors due to uncertainties in the nuclear charge distribution.
By making small adjustments to the assumed values for the half-density radius, $c$, and the skin-thickness, $t$ [Eq.~(3) of the main text], we gauge the sensitivity of the hyperfine constants to uncertainties in these parameters.
For small changes $\delta c$ and $\delta t$ around central values for $^{211}$Fr, we find $\delta \A/\A \simeq -0.075\, \delta c/c$, and $\delta \A/\A \simeq 0.0009\, \delta t/t$, as shown in Fig.~\ref{fig:dRrms}.
Therefore, errors stemming from uncertainties in the nuclear charge radii are negligible.
In terms of the root-mean-square radius, $r$, this implies $\delta \A/\A \simeq -0.086\, \delta r/r$, which can be used to estimate the relative Breit-Rosenthal corrections between neighboring isotopes.

\bibliography{hfs}

\end{document}